\documentclass[a4paper,20pt]{article}
    \usepackage[T1]{fontenc}
    \usepackage{graphicx}
    \usepackage{epsfig}
    \usepackage{amsmath}
    \usepackage{amsfonts}
    \renewcommand{\abstract}{}
\begin{document}
\makeatletter
\renewcommand{\@oddhead}{\textit{YSC'17 Proceedings of Contributed Papers} \hfil \textit{J. Dr\k{a}\.{z}kowska, M. Hanasz, K. Kowalik}}
\renewcommand{\@evenfoot}{\hfil \thepage \hfil}
\renewcommand{\@oddfoot}{\hfil \thepage \hfil}
\fontsize{11}{11} \selectfont

\title{Particle module of Piernik MHD code}
\author{\textsl{J. Dr\k{a}\.{z}kowska$^{1}$, M. Hanasz$^{1}$, K. Kowalik$^{1}$}}
\date{}
\maketitle
\begin{center} {\small $^{1}$Toru\'{n} Centre for Astronomy, Nicolaus Copernicus University, Toru\'{n}, Poland\\
drazkowska@astri.umk.pl}
\end{center}

\begin{abstract}
Piernik is a multi-fluid grid magnetohydrodynamic (MHD) code based on the Relaxing Total Variation Diminishing (RTVD) conservative scheme. The original code has been extended by addition of dust described within the particle approximation. The dust is now described as a system of interacting particles. The~particles can interact with gas, which is described as a fluid. The comparison between the test problem results and the results coming from fluid simulations made with Piernik code shows the most important differences between fluid and particle approximations used to describe dynamical evolution of dust under astrophysical conditions.
\end{abstract}

\section*{Introduction}
\indent \indent Planets are formed in disks surrounding young stars, which are called protoplanetary disks. The disks initially consist of gas and fine dust. The dust particles in disk collide because relative velocities are induced by random as well as systematic motions of grains in
the gaseous medium. During the collisions the small dust grains easily stick together due to mutual attractive forces like e.g. Van der Waals attraction \cite{dominik}. Details of this process can be examined in laboratory measurements. Experiments show us that the simple hit-and-stick collisions produce complicated fractal structures of agglomerates \cite{blumwurm}. Further growth of the grains depends on relative velocities of colliding particles. The agglomerates can still grow but can also be destroyed when the relative velocity during the collision is greater than some critical value. The laboratory experiments concerning this issue are highly complex and hard to perform. Nevertheless, their results give us useful information about possible ways of dust grains coupling into planetesimals that we can use in numerical simulations.

\section*{Piernik MHD code}
\indent \indent Piernik is a multi-fluid grid MHD code created and developed in Center for Astronomy of Nicolaus Copernicus University in Toru\'n. It is based on the Relaxing TVD conservative scheme presented by~\cite{jinxin} and~\cite{tracpen}. Piernik can be used to examine dynamics of ionized or neutral gas, as well as dust treated as a pressureless fluid. The code computes conservative fluid variables (fluid density, momentum, total energy density) for~each cell of~the~computational grid. The basic scheme has been extended by addition of many facilities which are useful in astrophysical fluid-dynamical simulations, e.g. cosmic rays influence, fluid interactions, Ohmic resistivity module and selfgravity module.
The current public version of Piernik code is available through the web-page {\it http:{/}{/}piernik.astri.umk.pl{/}}.
See \cite{piernik1,piernik2,piernik3,piernik4} for~a~more detailed description of~the~code.

\section*{Particle module of Piernik MHD code}
\indent \indent Dust can be described both in fluid and particle approximation. In the particle module of Piernik code the dust component is described as a system of identical independent particles that can interact with each other. The particles can also interact with gas, which is considered as a fluid and simulated with the~aid of~Piernik code. The code computes position and velocity of~every particle during each time step. The~equations of~motion are solved using fast and simple numerical scheme known as Verlet leap-frog method \cite{hockney}.

\begin{figure}[t]
\begin{minipage}[t]{.95\linewidth}
\centering
\epsfig{file = 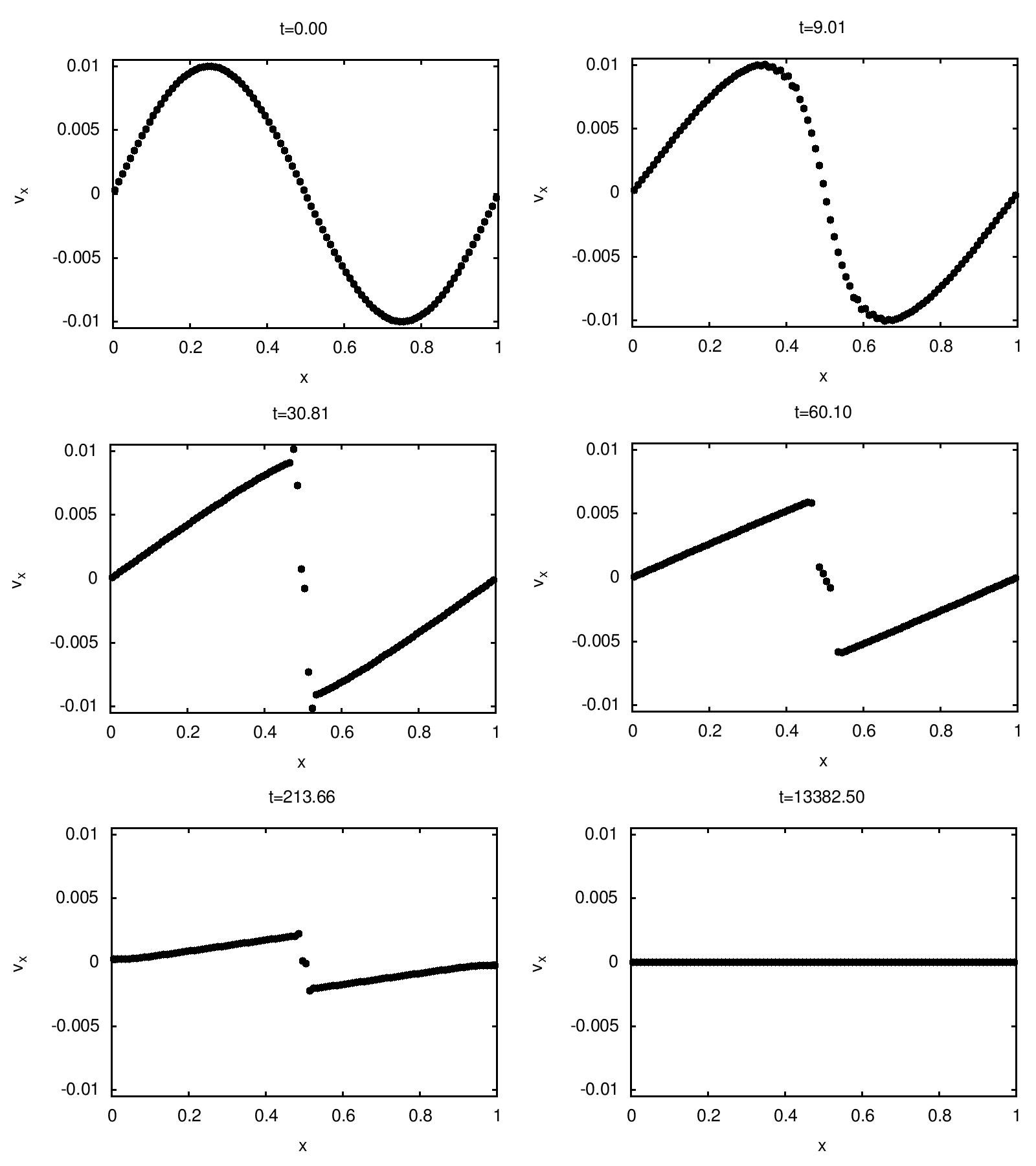, width = .45\linewidth}
\hfill
\includegraphics[width = .45\linewidth]{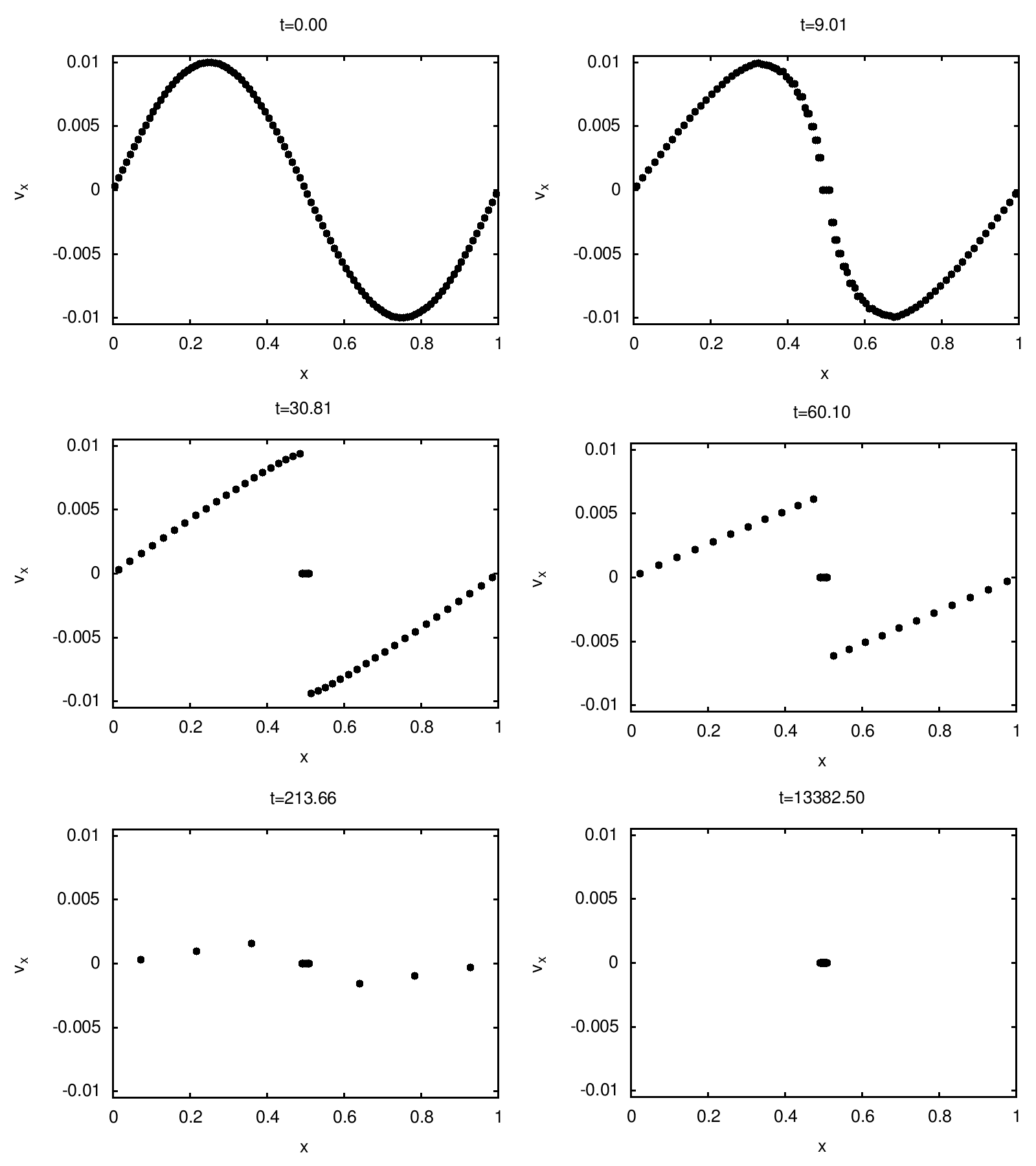}
\caption{Results of 1D sinusoidal wave test obtained for pure dust component in fluid (left panel) and particle (right panel) approximation respectively. For the fluid approximation result, each dot on the graphs is the value of velocity for a single cell of the grid. In the graphs displaying result given by the particle approximation, each dot is the value of velocity for a~single dust particle. The apparent difference between the results obtained with both of these methods is caused by the fact that in the particle simulations the values of physical quantities are specified only in the particles locations, whereas in the fluid simulations all the physical quantities are computed for every cell of the domain, even if density is very small.}
\end{minipage}
\end{figure}

\begin{figure}[!ht]
\begin{minipage}[t]{.95\linewidth}
\centering
\epsfig{file = 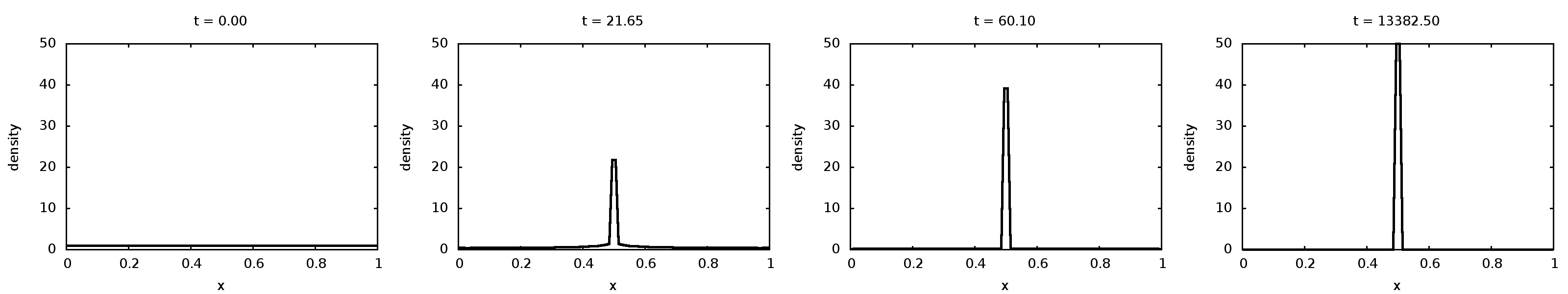, width = .95\linewidth}
\hfill
\caption{Density profile evolution obtained in fluid simulation of 1D sinusoidal wave. Initially the density distribution is uniform. Evolution of the system leads to the progressive enhancement of the central density peak. At the end of the simulation the fluid density profile is represented by one peak of density, which is equivalent of the cluster of particles obtained in particle simulation in the center of computational domain (right panel of Figure~1).}
\end{minipage}
\end{figure}

\section*{Results}
\indent \indent The particle module of Piernik code has been carefully tested. To compare fluid and particle approximations applied for the dust component we carried out several test problem simulations with the same initial conditions applied in both approaches.

The first test problem relies on the analysis of 1D sinusoidal velocity perturbation of the pure homogeneous dust component. The fluid approximation result, which can be verified by an analytical solution of the Burger's equation, displays a conversion of the initial sinusoidal velocity profile into the sawtooth profile and then smoothing until the flat profile, as~shown in~the~left panel of~Figure~1. The discontinuity in the velocity profile corresponds to~a~shock front.

In the absence of interactions the particle model leads to multiple velocity values in the velocity profile. To avoid the unphysical evolution of the particle system we have introduced interactions between particles analogous to inelastic collisions. We tested two variants of such an interactions. In the first case the particles stick when they meet each other in the same grid cell. In the second case two particles stick when distance between them is smaller than some specified radius corresponding to the radius of single particle. In both cases velocity of the resulting aggregate is center of mass velocity of the interacting particles before the collision. Results obtained in both cases are qualitatively similar if the sticking radius used in the second method corresponds to the size of single grid cell.

The result coming from the particle approach with the interactions taken into account (right panel of~Figure~1) still appears different than the result given by the fluid approximation, because the particles group together into clusters.
In the fluid simulations all the physical quantities are computed for every cell of the domain, even if density is very small. In the particle simulations the values of physical quantities are specified only in the particles locations. The fluid density profile at the end of the fluid simulation is represented by one peak of density (Figure~2). Respectively, at the end of 1D particle simulation all the particles are grouped together into one aggregate.

The differences between the fluid and particle approaches are much more significant in 2D tests. We~present results of two identical dust fronts collision test. At the beginning of the fluid simulation fluid density across both of these fronts is homogeneous. For particle approach random initial particles positions across the fronts have been chosen. The left front moves initially two times faster than the right one. The result of fluid simulation is shown in left panel of Figure 3. The fronts merge and move together with the velocity resulting from the momentum conservation law. On a longer timescale the front diffuses over the whole computational domain.

The results coming from particle simulation differ significantly (right panel of Figure 3). The fronts merge for short time and then they come apart. This fragmentation is probably caused by numerical coagulation of particles in the finite-size grid.

\begin{figure}[t]
\begin{minipage}[t]{.95\linewidth}
\centering
\epsfig{file = 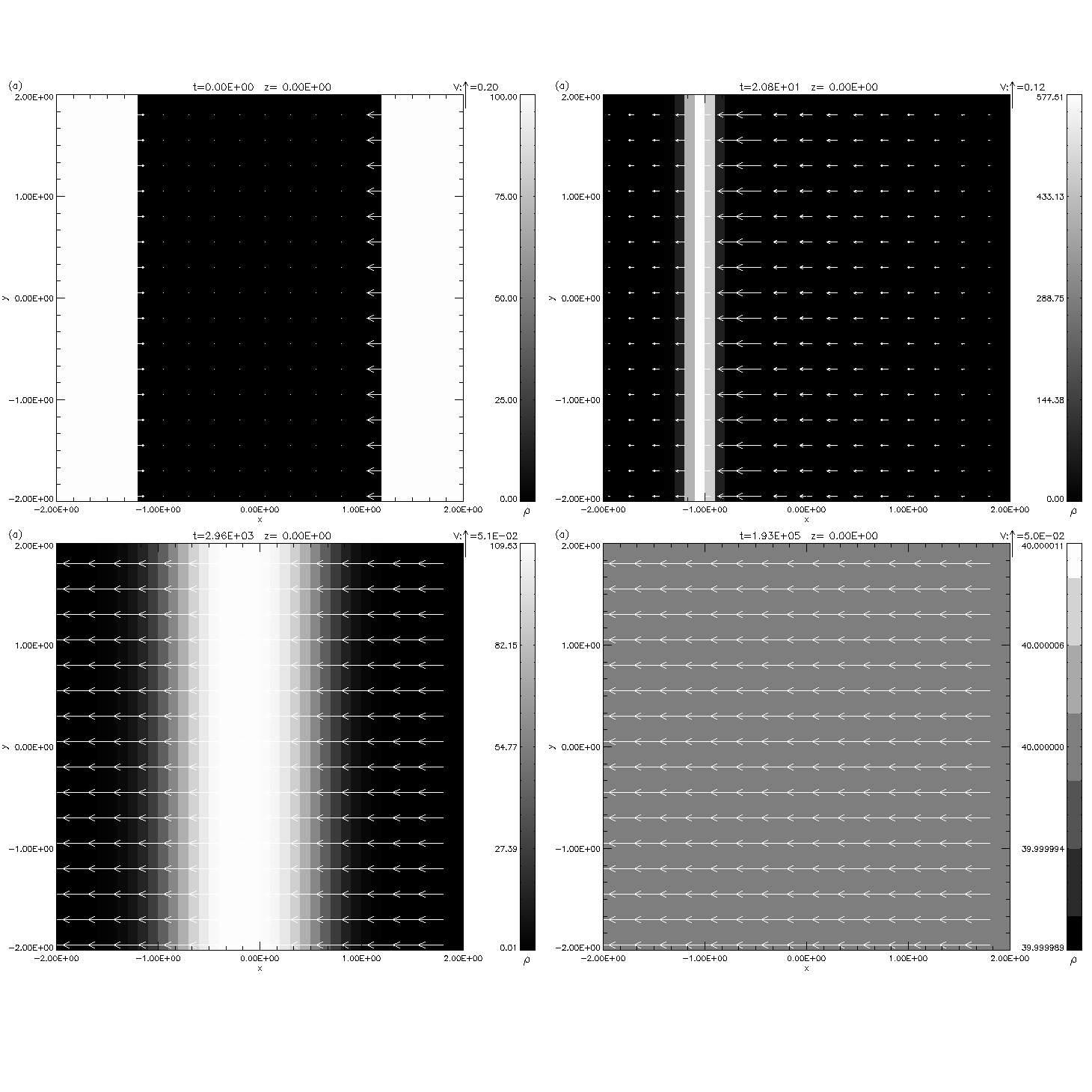, width = .45\linewidth}
\hfill
\includegraphics[width = .45\linewidth]{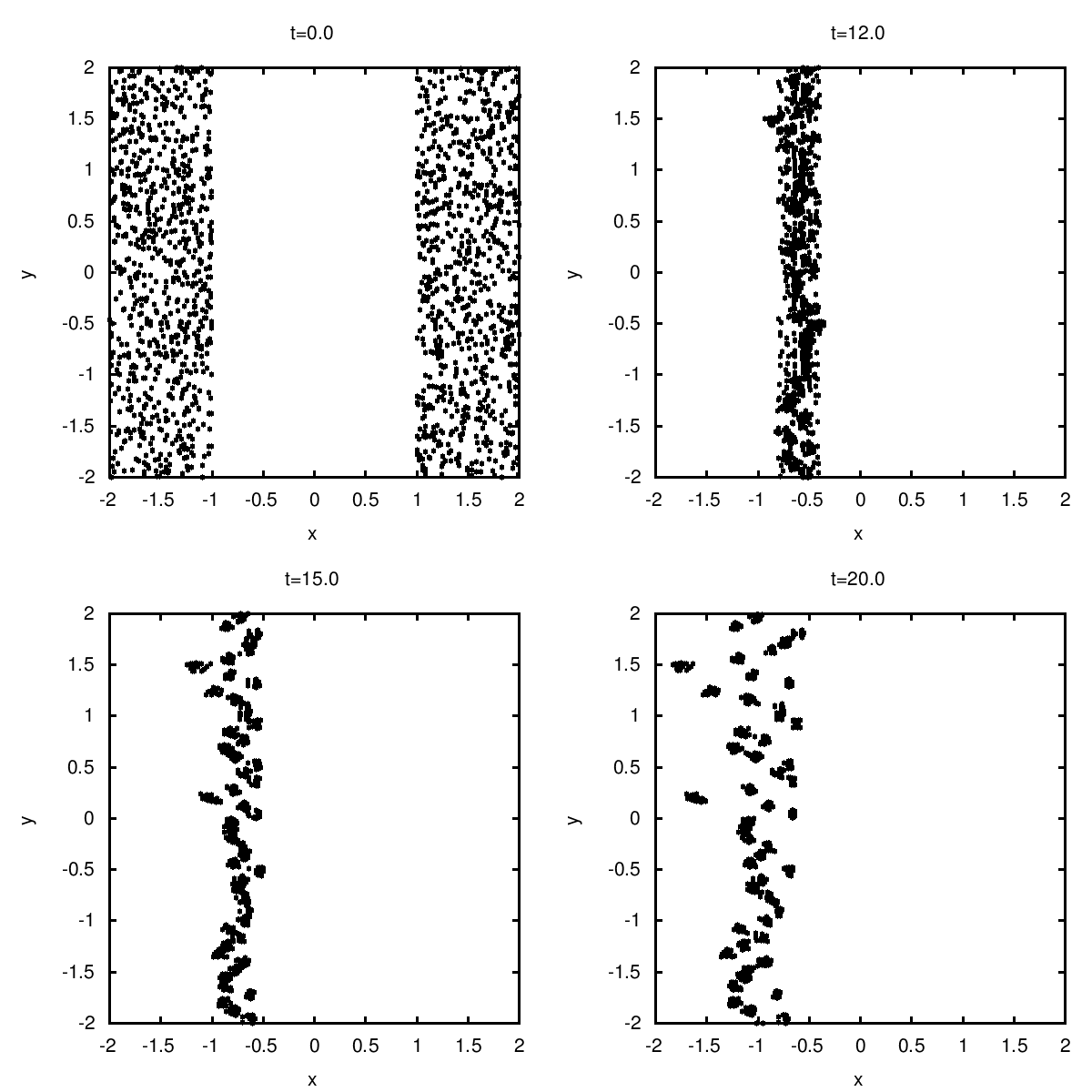}
\caption{Results of 2D dust fronts collision test obtained in fluid (left panel) and particle (right panel) approximation respectively. Density profiles are displayed for the fluid simulation. The lighter shade means the higher fluid density. For particle simulation result each dot marks the position of a single dust particle. The difference between the results obtained in fluid and particle approach is due to numerical coagulation of particles in the finite-size grid cells.}
\end{minipage}
\end{figure}

The initial condition of the last test problem relied on random positions and velocities embedded in gas of uniform density and velocity in 2D computational domain. We assume aerodynamic drag acting on particles and no back-reaction of particles upon the gas component. The simulation was made only in~particle approximation. Results obtained in the simulation are shown in Figure~4. The particles first form agglomerates and then move upward dragged by the gas. 
The agglomeration caused by hit-and-stick collisions is the first step towards planet formation. One can clearly see that significant number of the produced agglomerates have a~fractal structure, which is consistent with laboratory experiments results \cite{dominik}.

\begin{figure}[t]
\begin{minipage}[t]{.95\linewidth}
\centering
\epsfig{file = 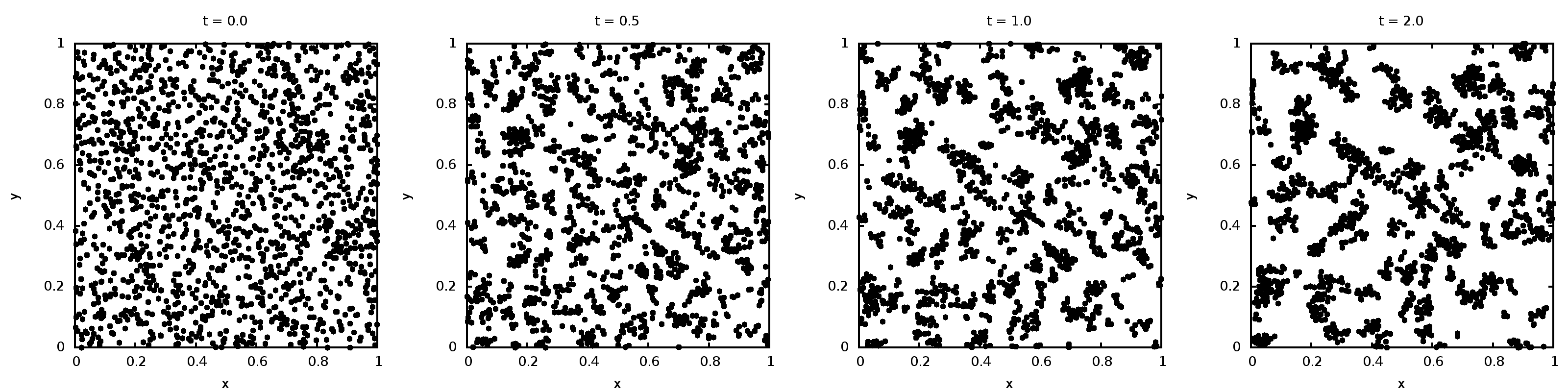, width = .95\linewidth}
\hfill
\caption{Results of 2D particle simulation with initial random particles positions and velocities distribution. Each dot marks the position of a single dust particle. This computational domain can be treated as a very small sector of a protoplanetary disk. The agglomeration caused by hit-and-stick collisions is the first step towards planet formation.}
\end{minipage}
\end{figure}

\section*{Summary} 
\indent \indent The dust in astrophysical simulations can be treated as a fluid or as a system of particles. These two approaches are not equivalent. Even the simple tests reveal differences between these two approaches.

The best way to make a complex simulations of dust evolution seems to be connection of both these approaches. We can use the fluid approximation to simulate the gas component and particle approximation for the dust component.
Implementation of simple hit-and-stick collisions for dust component considered as a system of particles allows to investigate only the very first stage of planetesimals formation.
We plan to improve the particle module of Piernik MHD code by implementation statistical Monte Carlo method with representative particles approach~\cite{mc}.

\section*{Acknowledgments}
\indent \indent This work was partially supported by Polish Ministry of Science and Higher Education through the grant 92/N-ASTROSIM/2008/0. The computations were performed on the HYDRA cluster at Toru\'n Centre for Astronomy.

\end{document}